\documentclass{nature} 
\usepackage[utf8]{inputenc}
\usepackage[english]{babel}
\usepackage{enumerate}
\usepackage{color}

\usepackage{dcolumn}
\usepackage{bm}
\usepackage{mathptmx}
\usepackage{pifont}
\usepackage{cancel}
\usepackage{amssymb}
\usepackage{float}
\usepackage{amsmath}
\usepackage{commath} 
\usepackage{physics}
\usepackage{upgreek}
\usepackage{mathrsfs}
\usepackage{lineno}
\usepackage{xcolor}
\usepackage{graphicx}
\usepackage[normalem]{ulem}
\newcommand\bluesout{\bgroup\markoverwith{\textcolor{blue}{\rule[0.5ex]{2pt}{1.1pt}}}\ULon}
\makeatletter
\let\saved@includegraphics\includegraphics
\AtBeginDocument{\let\includegraphics\saved@includegraphics}
\renewenvironment*{figure}{\@float{figure}}{\end@float}
\makeatother
\title{Intensity-driven Mott-like transition in a 2D Ruddlesden-Popper perovskite}

\author{Koushik Mondal,$^{1,2}$ Amit Haldar,$^{1,3}$ Indrajit Mondal,$^{3,4}$ Arpita Dutta,$^{1,3}$ Himanshu Bhatt,$^5$ Biswajit Mahata,$^{1,3}$ Soumili Banerjee,$^{1,3}$ Hirendra Nath Ghosh,$^{3,5,6}$ Satyaprasad P Senanayak,$^{3,4,6}$ and Shovon Pal$^{1,3,6}$}

\begin{document}
\maketitle
\begin{affiliations}
\item School of Physical Sciences, National Institute of Science Education and Research, Jatni, 752 050 Odisha, India
\item Department of Electrophysics, National Yang Ming Chiao Tung University, Hsinchu 300, Taiwan
\item Homi Bhabha National Institute, Training School Complex, Anushakti Nagar, 400 094 Mumbai, India
\item Nanoelectronic and Device Physics Lab, School of Physical Sciences, National Institute of Science Education and Research, Jatni, 752 050, Odisha, India.
\item School of Chemical Sciences, National Institute of Science Education and Research, Jatni, 752 050 Odisha, India
\item Center for Interdisciplinary Sciences, National Institute of Science Education and Research, Jatni, 752 050 Odisha, India
\end{affiliations}

\vspace{-20pt}

\begin{flushleft}
(\today)
\end{flushleft}

\begin{abstract}
Two-dimensional Ruddlesden-Popper-type halide perovskites have emerged as promising cutting-edge nonlinear optical materials due to their strong excitonic effects and carrier-density-dependent optical responses. In this work, we investigate the nonlinear optical properties and ultrafast carrier dynamics of (BA)$_2$PbI$_4$ under infrared excitation, resonant with the two-photon energy of its excitonic transition. Using a nonlinear femtosecond Z-scan measurement with a 1030\,nm pump pulse, we observe a pronounced intensity-dependent evolution of the nonlinear absorption coefficient and the refractive index. With increasing pump intensity, a switch from saturable absorption to reverse saturable absorption emerges, suggesting an intensity-driven Mott-like transition. To elucidate the underlying carrier dynamics, intensity-dependent transient absorption spectroscopy was performed under the same excitation intensities. This reveals a long-lived charge carrier population above the estimated Mott density, reflecting a clear phase transition from an exciton-dominated regime to a dense electron-hole plasma regime. Moreover, such an intensity-driven, high, and switchable nonlinear optical response, along with a Mott-like transition under infrared pulse excitation, makes the 2D (BA)$_2$PbI$_4$ perovskites suitable for all-optical operation in the infrared regime.
\end{abstract}

\maketitle


\section{Introduction}
Nonlinear optical materials with substantial third-order nonlinear optical responses and explicit intensity-dependent physical properties are in great demand for on-chip integration and all-optical device applications~\cite{Xu2020AM, Wang2021Small, Zhou2019AM, Kravtsov2018PRL}. Driven by the requirements of modern photonic applications, 2D materials, such as graphene and transition metal dichalcogenides (TMDCs), have been extensively explored. Graphene is claimed to have a robust and controllable nonlinear optical response because of its gapless electronic band structure, whereas TMDCs are recognized for their exciton-driven nonlinear response~\cite{Lan2020PRB, Wu2021SS, Mondal2024AOM, Kumar2009APL}. However, graphene exhibits significant broadband linear absorption owing to its gapless electronic structure, while the giant excitonic nonlinear response in TMDCs is largely confined to the monolayer limit, thereby restricting the optical interaction length and making the response strongly dependent on the number of layers. Preparation of such 2D materials in monolayer or even few-layer configurations, however, remains a challenge because it requires robust control of experimental conditions. In contrast, the layered 2D Ruddlesden-Popper-type halide perovskites (RPPs) of the form A$_2$BX$_4$ (where A is the organic spacer, B is the metal cation, and X is the halogen) have emerged as a cutting-edge material class in recent years. The layered structure comprises alternating inorganic perovskite sheets and organic spacer layers~\cite{Hong1992PRB, Posmyk2024AOM, Deng2020NatCom}. The organic layer has a larger bandgap and a lower dielectric constant as compared to the inorganic layer~\cite{Even2014CPC, Kitazawa2011JPCS, Abdelwahab2019AM}, resulting in strong spatial (quantum) and dielectric confinement effects. Unlike graphene and monolayer TMDCs, 2D RPPs combine strong excitonic interactions with comparatively large interaction lengths in readily fabricated solution-processed thin films, providing an attractive platform for achieving efficient nonlinear light-matter interactions without requiring atomically thin crystals. Quantum confinement in these materials leads to strong excitonic effects, which in turn enable significant linear and nonlinear optical properties. These include saturable absorption~\cite{Xu2020AM} and nonlinear optical absorption~\cite{Xu2020AM, Quan2019CR, Wang2019AOM} of intense coherent light, strong photoluminescence efficiency~\cite{Walters2015ACSN}. These properties have potential applications in optoelectronics and photonics, including multiphoton spectroscopy, infrared photodetectors, and higher-harmonic signal generation~\cite{Zhang2015ACSP, Lu2018AS, Saouma2017NatCom}.

In recent times, halide perovskite containing butyl-ammonium (BA) cations, i.e., (BA)$_2$PbI$_4$, a member of the same 2D-RPP family, has emerged as an excellent playground for nonlinear optical research, with robust excitonic effects at room temperature, similar to the 2D-TMDC materials. Notably, in these Ruddlesden-Popper-type quantum well structures, exciton-exciton interactions become significant, rendering the boson-like non-interacting-particle approximation invalid. As a result, excitons substantially contribute to the nonlinear optical response when driven by a strong external electromagnetic field~\cite{Abdelwahab2019AM, Hanamura1988MSE}. The large transition dipole moments, oscillation strengths, and binding energies~\cite{Blancon2018NatCom, Hansen2024AOM} of the electron-hole pairs (or excitons) in (BA)$_2$PbI$_4$ enable intense interaction with the external electromagnetic field, resulting in robust optical nonlinearity~\cite{Posmyk2024AOM}. The wavelength-dependent switching of nonlinear absorption, i.e., from saturable to reverse-saturable absorption in (BA)$_2$PbI$_4$ film has been demonstrated in the visible region of the spectrum, where the two-photon absorption coefficient is found to be $-2.4 \times 10^5$\,cm/GW~\cite{Liang2020OExp}, which is orders of magnitude higher than that of graphene ($-90$\,cm/GW)~\cite{Kumar2009APL}. (BA)$_2$PbI$_4$ is also known to exhibit a substantial third-order nonlinear optical susceptibility, $\chi^{(3)}$, and a robust third-harmonic generation efficiency when excited by an infrared pulse at the three-photon resonance of the 1$s$-exciton~\cite{Abdelwahab2018ACSN}. The strongly bound excitons and exciton-exciton interaction in (BA)$_2$PbI$_4$ further facilitate the density-driven electronic phase transition, namely the Mott-type insulator (exciton) to metal (plasma) transition~\cite{Abdelwahab2019AM}. The Mott transition occurs when tightly bound excitons dissociate into a dense electron-hole plasma, driven by the increased screening of the Coulomb interaction~\cite{Kappei2005PRL}. External perturbations, such as DC electric fields, temperature, and ultrafast light fields, can trigger the phase transition. Controlling the Mott transition with ultrashort pulses could enable reversible tuning of the material's nonlinear optical response~\cite{Zimmermann1988PSS}, paving the way for diverse infrared optoelectronic applications~\cite{Milloch2024npjS, Basov2017NM, Hao2023NatCom, Iwai2003PRL, Luibrand2024PRB, Lui2023AM}.

In this article, we exploit the robust third-order nonlinearity to drive the Mott-like transition in (BA)$_2$PbI$_4$ by systematically tuning the excitation intensity of a femtosecond infrared pulse at the two-photon resonance condition of the exciton energy. We identify a critical intensity $I_{\rm P}$, below which the system exhibits saturable absorption originating from the saturation of the two-photon absorption. In contrast, at higher excitation intensities, three-photon absorption takes over, resulting in a reverse-saturable absorption-type behavior. From the open-aperture Z-scan measurements, we obtain a two-photon absorption coefficient of $\approx -6.29 \times 10^6$\,cm/GW, which is higher than that of several 2D materials, such as graphene, TMDCs, and other perovskites. The closed-aperture Z-scan data show a remarkable intensity-dependent switching of nonlinear refraction from self-defocusing at low intensities to self-focusing at high intensities -- a direct evidence of the intensity-driven Mott-like transition. Intensity-dependent transient absorption experiments further reveal distinct charge-carrier dynamics above and below $I_{\rm P}$, thereby corroborating the dense electron-hole-plasma and excitonic phases, respectively.

\section{Result and Discussion}
\begin{figure}
    \centering
    \includegraphics[width=\linewidth]{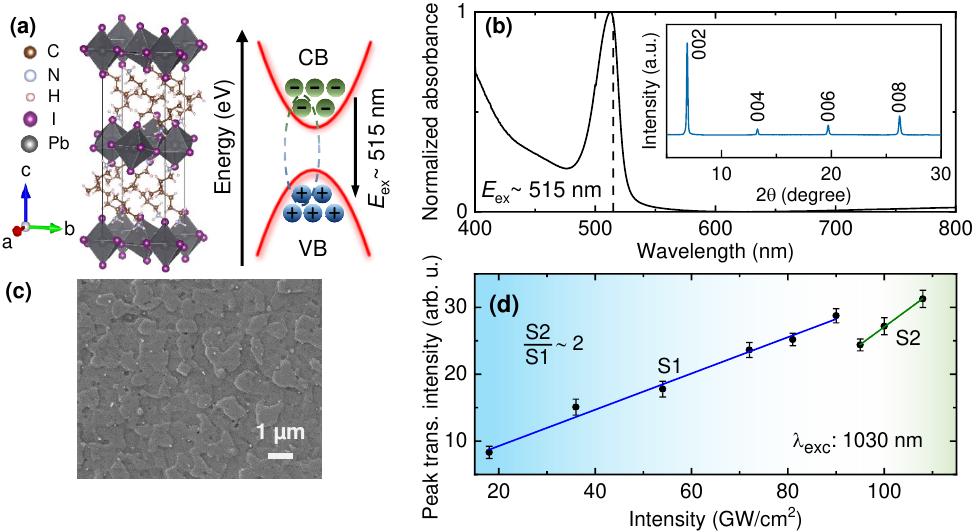}
    \caption{{\bf(a)} Schematic representation of the crystal structure and the excitonic band of (BA)$_2$PbI$_4$. {\bf(b)} Absorption spectrum of the as-prepared (BA)$_2$PbI$_4$ film showing a sharp 1$s$-excitonic peak at 515\,nm. The inset shows the XRD profile of the (BA)$_2$PbI$_4$ thin film on quartz glass substrate. {\bf(c)} SEM micrograph of (BA)$_2$PbI$_4$ thin film, showing a good surface coverage and film continuity. {\bf(d)} Variation of the transmitted peak intensity of a femtosecond light at 1030\,nm, which almost doubles above a critical intensity of $I_{\rm P}\approx90$\,GW/cm$^2$.}
    \label{fig1}
\end{figure}

Crystalline 2D (BA)$_2$PbI$_4$ samples are prepared using a solution processing method, which is detailed in the methods section. All experiments in the current investigation are conducted on thin films of the as-synthesized material that are spin-coated onto high-quality $z$-cut quartz substrates. The sample exhibits a typical Ruddlesden-Popper phase having two-dimensional perovskite-like slabs interleaved with cations. Figure~\ref{fig1}a shows a schematic of the structure with a single octahedral layer in the perovskite-like stack. The thickness of our films was measured to be approximately $0.18\pm0.01$\,$\mu$m, see Section~S1 of the Supporting Information. The thin-film XRD data presented in the inset of Figure~\ref{fig1}b show excellent agreement with earlier reported diffraction patterns~\cite{Kumar2026small, Cao2015JACS}, strongly supporting the phase formation, crystallographic orientation, and structural quality of the prepared (BA)$_2$PbI$_4$ thin films.

The absorption spectrum of the as-prepared (BA)$_2$PbI$_4$ film, shown in Figure~\ref{fig1}b, exhibits a sharp peak at $515$\,nm, corresponding to the 1$s$-excitonic transition (shown schematically in Figure~\ref{fig1}a) and agrees well with the reported literature values~\cite{Abdelwahab2019AM, Du2023Nano, Mathew2023AOM}. We note that the direct bandgap of the (BA)$_2$PbI$_4$ film, obtained from the Tauc plot analysis of the absorption data, is $2.376$\,eV (see Sections~S2 and~S3 of the Supporting Information), which is in good agreement with the literature values~\cite{Liang2020OExp}. The SEM micrograph in Figure~\ref{fig1}c further reveals a continuous and densely-packed polycrystalline morphology with well-defined grains, indicating the formation of a high-quality phase-pure film on the substrate. Remarkably, the sample shows a distinct intensity-dependent transmittance of femtosecond light at 1030\,nm, which corresponds to the two-photon absorption associated with the excitonic transition; see Figure~\ref{fig1}d. Specifically, we identify a critical intensity, $I_{\rm P}\approx90$\,GW/cm$^2$, across which the variation of transmittance with the incident intensity almost doubles. This suggests that our material exhibits two distinct phases with unique carrier dynamics, which can be systematically controlled by varying the intensity of two-photon excitation.

The third-order nonlinear optical response of (BA)$_2$PbI$_4$ film is investigated using the nonlinear Z-scan technique~\cite{Abdelwahab2019AM}. Z-scan measurements in the open-aperture (OA) and closed-aperture (CA) modes allow us to determine the sign and magnitude of the nonlinear absorption coefficient and the nonlinear refractive index, respectively. A focused femtosecond (fs) laser beam (with a wavelength of 1030\,nm, pulse duration of 370\,fs, and a repetition rate of 100\,kHz) is used to pump the (BA)$_2$PbI$_4$ film that is placed on a linear, motorized translation stage. The pump intensity $I_0$ was systematically varied in the range $18$\,GW/cm$^2$ to $125$\,GW/cm$^2$. As mentioned earlier, the 1030\,nm excitation was intentionally chosen since it lies in two-photon resonance with the 1$s$-excitonic transition of the as-prepared (BA)$_2$PbI$_4$ film. The diameter of the beam waist is 31\,$\mu$m, and the Rayleigh length ($Z_0$) is calculated to be 3\,mm, larger than the thickness of the (BA)$_2$PbI$_4$ film. When the Rayleigh length exceeds the film thickness, the beam spot size remains uniform throughout the sample thickness. This validates that the beam interacts with the sample at a nearly constant intensity, yielding more reliable and interpretable data.

\begin{figure}
    \centering
    \includegraphics[width=\linewidth]{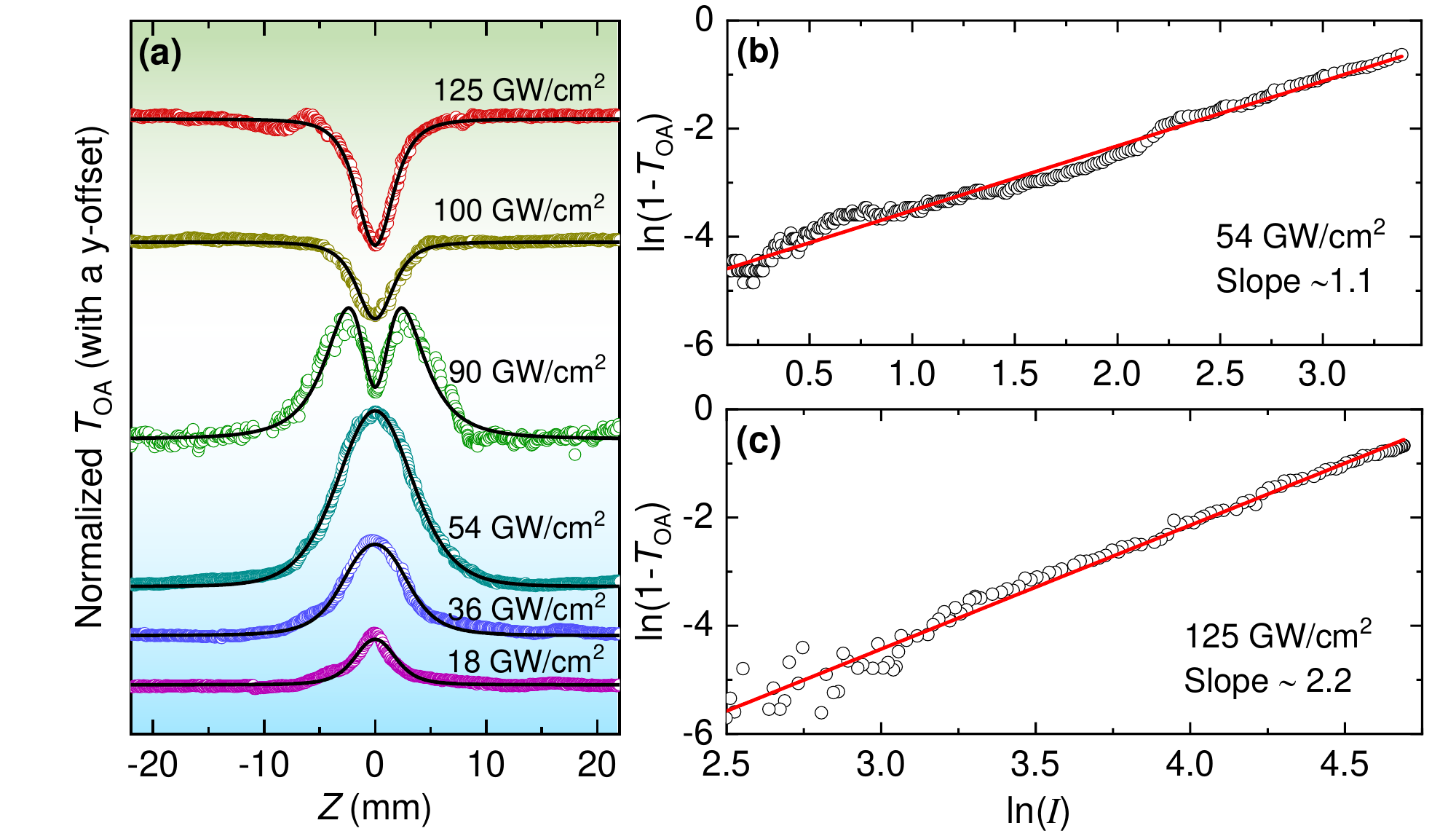}
    \caption{{\bf(a)} Normalized open-aperture (OA) Z-scan transmittance data of (BA)$_2$PbI$_4$ film under the excitation of 1030\,nm, 370\,fs laser beam at different on-axis peak intensities. The black solid lines represent the theoretical fitting to the experimental data using the model as described in Section~S5 of the Supporting Information. Linear variation of $ln|1-T_{\rm OA}|$ with $ln(I)$ under the excitation intensities of {\bf(b)} 54\,GW/cm$^2$ and {\bf(c)} 125\,GW/cm$^2$, displaying a slope of 1.1 (indicating two-photon absorption) and 2.2 (indicating three-photon absorption), respectively.}
    \label{fig2}
\end{figure}

In the OA mode of the Z-scan experiments, the sample transmittance is measured with the entire laser beam incident on the detector, without any aperture in front of the detector. Figure~\ref{fig2}a shows the OA transmittance data ($T_{\rm OA}$) at different excitation intensities $I_0$. We find that for $I_0<I_{\rm P}$, the normalized OA transmittance gradually increases as the sample approaches the focal point (i.e., $Z=0$), indicating that saturable absorption (SA) is the dominating phenomenon. At $I_0=I_{\rm P}\approx90$\,GW/cm$^2$, the OA transmittance shows a dip at $Z=0$, signaling a reverse saturable absorption (RSA) process, where the nonlinear optical response becomes RSA-in-SA type. For $I_0>I_{\rm P}$, the SA effect completely disappears, and only the RSA effect dominates the nonlinear optical response of the (BA)$_2$PbI$_4$ film. To identify the nature of the nonlinear phenomena involved in such an intriguing intensity-dependent nonlinear optical response, we further examine the OA transmittance data by plotting $ln|1-T_{\rm OA}|$ against $ln(I)$, shown in Figure~\ref{fig2}b and~\ref{fig2}c for two different intensities. Here, $I=I_0/(1+Z^2/Z^2_0)$, represents the position-dependent intensity along the Z-axis. We find that when $I_0<I_{\rm P}$, the $ln|1-T_{\rm OA}|$ varies linearly with $ln(I)$ with a slope of 1.1, which indicates the presence of two-photon absorption~\cite{Mondal2021OptLett}. In contrast, for $I_0>I_{\rm P}$, the slope increases to 2.2, indicating three-photon absorption~\cite{Mondal2021OptLett}. A detailed analysis of the OA Z-scan transmittance data is presented in Section~S5 of the Supporting Information. This remarkably corroborates our observations in Figure~\ref{fig1}d. We thus conclude that in the low-intensity regime, saturation of two-photon absorption leads to an SA-type response. In the high-intensity regime, however, the three-photon absorption causes a strong RSA-type response.

We find that the (BA)$_2$PbI$_4$ film shows an exceptionally high two-photon absorption coefficient reaching values of $-6.29\times 10^6$\,cm/GW at $I_0 = 90$\,GW/cm$^2$ (a detailed comparative analysis of the reported absorption coefficients for different material systems is provided in Table 3 of the Supporting Information). Previously, it has been shown that 2D-(BA)$_2$PbI$_4$ exhibits an anomalous dispersion in the real part of the dielectric constant near its excitonic transition energy~\cite{Abdelwahab2019AM}. In our work, the 1030\,nm pump is resonant with the two-photon excitonic transition in (BA)$_2$PbI$_4$, resulting in an efficient carrier generation via two-photon absorption in the low-pump-intensity regime, i.e., for $I_0<I_{\rm P}$. The increased excitonic density thereafter enhances the SA behavior of the (BA)$_2$PbI$_4$ film. With sufficiently strong excitation intensity (i.e., for $I_0>I_{\rm P}$), strong dipolar screening and phase-space filling set in, leading to a renormalization (reduction) of the excitonic bandgap -- a phenomenon that concomitantly drives the Mott-like transition to an optically-induced dense electron-hole plasma phase~\cite{Abdelwahab2019AM, Gong2024NC}. In this regime, as the photoexcited carrier density approaches the critical Mott density, exciton ionization induced by Coulomb screening produces a dense electron-hole plasma (free carrier). These free carriers introduce additional absorption channels through free-carrier absorption, thereby reinforcing the observed reverse saturable absorption, and the response becomes three-photon absorption.

\begin{figure}[t!]
    \centering
    \includegraphics[width=0.45\linewidth]{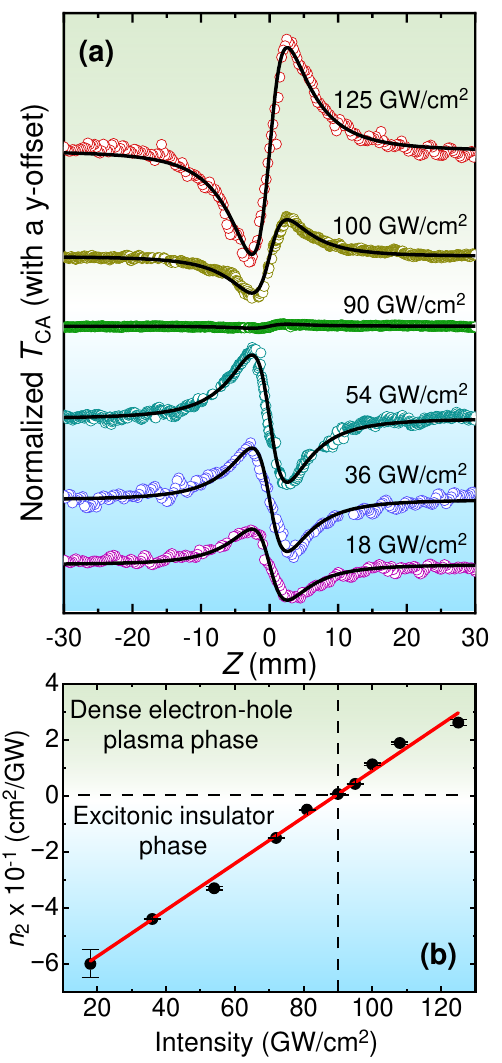}
    \caption{{\bf(a)} Normalized closed-aperture (CA) Z-scan transmittance data of (BA)$_2$PbI$_4$ film under the excitation of 1030\,nm, 370\,fs laser beam at different on-axis peak intensities. The black solid line represents the theoretical fitting to the experimental data, as described in Section~S6 of the Supporting Information. {\bf(b)} Variation of the nonlinear optical refractive index ($n_2$) at 1030\,nm with the on-axis pump intensity ($I_0$).}
    \label{fig3}
\end{figure}

Figure~\ref{fig3}a shows the normalized closed-aperture (CA) transmission data under different incident pump intensities. Upon normalizing the CA response with respect to the OA response, contributions from the nonlinear absorption can be removed, and information on the pure nonlinear refraction is obtained. Note that the CA Z-scan data (see Figure~\ref{fig3}a) show highly symmetric peak-valley (or a valley-peak) profiles over the entire excitation intensity range. Such a symmetric response is characteristic of an instantaneous electronic nonlinearity and indicates that the contribution from cumulative thermal effects is negligible. In general, cumulative thermal lensing typically introduces an asymmetric distortion in the Z-scan trace because the sample retains heat as it translates through the focal region, leading to a dragging or flattening of one side of the peak or valley. The absence of such asymmetry in our measurements, therefore, rules out any significant thermal contribution. Even in the case of the highest pump intensity excitation, the peak-to-valley difference $(\Delta Z_{\rm p-v})$ was 4.9, which is well below the thermal lensing regime of 5.1 ($=1.7\times Z_0$) as per our experimental conditions~\cite{Sheik1990IEEE,Seetharaman2017JPCC}.

Interestingly, we also find that our (BA)$_2$PbI$_4$ films show negative $n_2$ or self-defocusing type of nonlinear refraction for intensities $I_0<I_{\rm P}$. For $I_0>I_{\rm P}$, we observe a transition of $n_2$ to positive values, indicating a self-focusing type of nonlinear refraction, as shown in Figure~\ref{fig3}b. This transition of nonlinear refractive index is characteristic of the Mott-like transition that has been reported earlier using wavelength tuning~\cite{Abdelwahab2019AM}. In our experiment, for the low-pump-intensity regime ($I_0<I_{\rm P}$), the photoexcited carrier density remains below the critical Mott density, and the optical response is dominated by the strongly-bound excitons. Since the excitation wavelength of 1030\,nm is in two-photon resonance with the 1s excitonic transition, the nonlinear susceptibility is primarily governed by bound excitons. The generated exciton population partially bleaches the excitonic absorption through phase-space filling, reducing the oscillator strength of the excitonic transition. Now, in accordance with the Kramers-Kronig relation, any modification of the absorption spectrum must be accompanied by a corresponding change in the refractive index~\cite{Boyd2020AP}. Consequently, the excitonic bleaching produces the observed negative nonlinear refractive index (self-defocusing) in the low-intensity regime. This behavior is consistent with the well-established theory of resonant semiconductor nonlinear optics, in which the refractive response arises from the dispersive counterpart of the excitonic absorption spectrum~\cite{Abdelwahab2019AM}. However, in the high-pump-intensity regime, i.e., for $I_0>I_{\rm P}$, the photoexcited carrier density approaches the critical Mott density. As a result, enhanced Coulomb screening progressively weakens the electron-hole attraction, while phase-space filling further suppresses the excitonic oscillator strength. Simultaneously, many-body interactions give rise to band gap renormalization, leading to a redshift and broadening of the excitonic resonance. These processes substantially modify the spectral dependence of the dielectric function. Once excitons ionize into a dense electron-hole plasma (free carriers), this ionized plasma further contributes to the refractive response through carrier-induced changes in the dielectric function. The combined effects of exciton bleaching, Coulomb screening, bandgap renormalization, and free-carrier refraction alter the dispersive response sufficiently to reverse the sign of the nonlinear refractive index, resulting in the experimentally observed transition from self-defocusing to self-focusing~\cite{Haug1985JOSA,Price2015NC}.

Additionally, the extracted nonlinear refractive index ($n_2$) varies linearly with the incident laser intensity, which is a characteristic signature of a third-order electronic nonlinear process. In contrast, cumulative thermal effects generally exhibit a nonlinear (often quadratic or higher-order) dependence on the excitation power due to progressive (time-dependent) heat accumulation and the associated changes in the local refractive index. The observed linear scaling of $n_2$, therefore, further confirms the electronic origin of the measured nonlinearity. Such a linear dependence is also observed in the transient absorption signal over the same pump-fluence range (as discussed later), thereby substantiating the dominance of electronic nonlinearity.

\begin{figure}[t!]
    \centering
    \includegraphics[width=\linewidth]{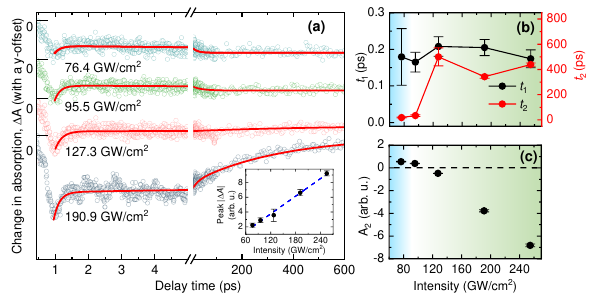}
    \caption{{\bf(a)} Temporal dynamics of the two-photon transient absorption bleach signal at 515\,nm in (BA)$_2$PbI$_4$ system at different pump intensities for a pump wavelength of 1030\,nm. The red-solid lines correspond to the bi-exponential fit used to model the two decay rates. The inset shows the magnitude of $\Delta A$ at different pump intensities. The blue dashed line represents the linear dependence of the absorption amplitudes on intensity, implying a third-order nonlinear process. {\bf(b)} The variation of $t_1$ (fast decay) and $t_2$ (slow decay). $t_1$ represents the fast time dynamics corresponding to the hot electron relaxation, and $t_2$ represents relaxation dynamics of the exciton that undergoes a dramatic change when the system evolves from the excitonic phase to the dense electron-hole plasma phase at the onset of the critical pump intensity ($I_{\rm P}$), identified from the nonlinear Z-scan measurement. {\bf(c)} The amplitude of the slow exponential decay, $A_2$, changes sign below $I_{\rm {P}}$, which corroborates the corresponding sign reversal of the nonlinear refractive index across the phase transition.}
    \label{fig4}
\end{figure}

To scrutinize the intensity-driven carrier dynamics in (BA)$_2$PbI$_4$ across the Mott-like transition, we performed transient absorption spectroscopy using a 1030\,nm pump pulse. The time transients of the two-photon absorption bleaching signal at the excitonic band edge (515\,nm) for different pump intensities are shown in Figure~\ref{fig4}a. The time dynamics evidently show two decay components. The fast component (with a time constant $t_1$) is attributed to the relaxation of the hot electrons, which remains nearly unchanged as a function of pump intensity. In contrast, the slow component (with a time constant $t_2$) exhibits a pronounced increase from a few tens of picoseconds to several hundreds of picoseconds as the pump intensity is tuned across $I_{\rm P}$, shown in Figure~\ref{fig4}b. At low pump intensities (i.e., $I_0<I_{\rm P}$), the $t_2$ lies on the order of exciton relaxation timescales. For pump intensities where the excited carrier density is above the estimated Mott density (i.e., $I_0>I_{\rm P}$), which is roughly 9$\times$10$^{18}$\,cm$^{-3}$ for (BA)$_2$PbI$_4$~\cite{Mott1990,Ghosh2026AOM}, the excitons get ionized into a dense electron-hole plasma due to the enhanced Coulomb screening. This would lead to a suppression of the non-radiative recombination channels, leading to a long-lived population of free charge carriers with an increased $t_2$. Note that the peak absorption amplitude of the transient absorption bleach signal, i.e., $|\Delta A|$, is linearly dependent on the incident pump intensity; see the inset of Figure~\ref{fig4}a. This indicates that the amplitude of the measured bleach signal is directly proportional to the population of photo-excited carriers. Since the carrier generation itself arises from a nonlinear two-photon absorption process associated with the excitonic transition, the linear dependence further corroborates third-order nonlinearity and a very high two-photon absorption coefficient. Note that there is an additional long decay time component ($>1$\,ns) that is assigned to radiative/non-radiative defect mediated recombination~\cite{Zheng2018JMCA, Magill2022AIPA}. We find that the variation of the amplitude $A_2$ corresponding to the slow exponential decay, as shown in Figure~\ref{fig4}c, changes sign in the optically-induced dense electron-hole plasma phase as compared to the excitonic insulator phase, corroborating the variation of $n_2$ in Figure~\ref{fig3}b. Our observations further substantiate that the on-axis phase shift (used to extract $n_2$) associated with absorption at 1030\,nm is correlated with the sign of the excitonic absorption amplitude. 
 
\section{Conclusion}
In conclusion, we demonstrate that a 2D-(BA)$_2$PbI$_4$, which is a Ruddlesden-Popper perovskite, exhibits a remarkably strong room-temperature, intensity-dependent third-order nonlinear optical response under femtosecond pulsed excitation in the near infrared region. The material shows pronounced nonlinear absorption and a change in refractive index when excited with a 1030\,nm femtosecond pulse under two-photon resonance with the excitonic transition at 515\,nm. Upon increasing the pump intensity, we observe a clear switching in the optical response, in which a saturable absorption-type behavior changes to a reverse saturable absorption behavior, with a concomitant reversal of the sign of the nonlinear refractive index. Our observations clearly reflect an intensity-driven Mott-like transition from an exciton-dominated regime to an optically-induced dense electron-hole plasma phase. The temporal dynamics of the charge carriers hint at a long-lived population of free charge carriers, observed when the pump intensity exceeds the estimated Mott density, indicating the ionization of excitons into the electron-hole plasma due to strong Coulomb screening. Our findings suggest that (BA)$_2$PbI$_4$ exhibits a very high two-photon absorption coefficient and can act as a promising platform for nonlinear photonics and optoelectronics in the infrared spectrum.      

\clearpage

\clearpage

\clearpage

\clearpage

\clearpage
\section*{Experimental Methods}
{\bf Sample preparation.} Thin films of (BA)$_2$PbI$_4$ are prepared using n-butylammonium iodide (BAI, obtained from Great Cell Solar Ltd.) and lead(II) iodide (PbI$_2$, obtained from TCI) as precursor materials, both used without further purification. The precursor solution is formulated by dissolving BAI (0.25\,M) and PbI$_2$ (0.25\,M) in a mixed solvent system of N,N-dimethylformamide (DMF) and dimethyl sulfoxide (DMSO) with a volume ratio of 3:1. Before film deposition, $z$-cut quartz substrates are sequentially cleaned by ultrasonication and subsequently treated with UV-ozone for 15\,mins to ensure a contaminant-free hydrophilic surface. A total volume of 40\,$\mu$L of the precursor solution is dispensed onto the pre-cleaned substrates, and spin-coated at 5000\,rpm for 60\,s. During the spin-coating process, 90\,$\mu$L of chlorobenzene is dynamically applied as an anti-solvent, 8\,s after the onset of spinning to promote rapid crystallization and uniform film formation. Following deposition, the films are thermally annealed at 100\,$^{\circ}$C for 30\,min to achieve complete phase formation and improved crystallinity. All steps related to precursor preparation and film fabrication are carried out inside a nitrogen-filled glove-box, maintaining oxygen and moisture levels below 0.1\,ppm to prevent ambient degradation.

\noindent{\bf Z-scan measurements.} The Z-scan, a single-beam technique, is employed to evaluate the nonlinear optical properties of (BA)$_2$PbI$_4$. In this method, the sample is translated in the $Z$-direction, i.e., along the axis of a focused Gaussian beam, and the far-field intensity is measured as a function of the sample position. We have used a 370\,fs laser pulse, operating at a 100 kHz repetition rate and centered at 1030\,nm, to excite the material. The pulsed laser beam is focused to a spot size of 31\,$\mu$m using a lens with a focal length of 150\,mm. The as-prepared (BA)$_2$PbI$_4$ film is translated from $-Z$ to $+Z$ position (through the focus at $Z=0$) along the beam propagation direction using a motorized translation stage. In open-aperture mode, the entire transmitted beam is measured, which yields the nonlinear absorption coefficients. In contrast, in the closed-aperture mode, the nonlinear refractive index is determined by measuring the partially transmitted beam through an aperture.

\noindent{{\bf Pump-probe experiments.}} For the transient absorption spectroscopy, a Ti:sapphire amplified laser system is used to generate femtosecond laser pulses of 800\,nm operational wavelength, 5\,mJ pulse energy, 100\,fs pulse width, and 1\,kHz repetition rate. The laser output is split into two beams (pump and probe) using a 70:30 beam splitter. An optical parametric amplifier is then used to generate the pump excitation wavelength of 1030\,nm. To acquire the transient absorption spectra, a Helios pump-probe spectrometer is used. The probe is delayed using a mechanical delay stage. The 800\,nm monochromatic light beam is focused onto a sapphire crystal to generate a white-light continuum probe beam spanning 420\,nm to 780\,nm. Both the visible probe pulse and the pump beams overlap at the sample position. A CMOS detector is used to detect the transmitted probe. The difference in the absorption of the probe pulse ($\Delta A$) in the presence and absence of the pump ($A_{\rm pump\,ON}-A_{\rm pump\,OFF}$) is obtained for the range of wavelengths ($420-780$\,nm) and delay time (100\,fs -- 7\,ns).

\begin{addendum}
\item[Author Contributions] All authors contributed to the discussion and interpretation of the experiment and to the completion of the manuscript. K.M. and A.H. performed and analyzed the Z-scan measurements. I.M., under the supervision of S.P.S., prepared and characterized the samples. H.B., under the supervision of H.N.G., performed the transient absorption and time-resolved luminescence spectroscopies. H.B., B.M., and A.D., under the supervision of S.P., analyzed the transient absorption data. B.M. and S.B. performed and analyzed the linear transmission experiments. K.M. and S.P. conceived the project, while S.P.S., and S.P. supervised the experiments. K.M., A.D., and S.P. drafted the manuscript with inputs from all authors.

\item[Acknowledgments] A.H., I.M., A.D., S.P.S., and S.P. acknowledge the support from DAE through project Nos. RIN4001 and RNI4011. K.M. acknowledges research support from NYCU and NSTC, Taiwan, under Grant: NSTC114-2811-M-A49-551. H.B. acknowledges NISER for the research fellowship. H.N.G. acknowledges the support from DAE through project Nos. RIN4002 and RNI4011, and is grateful to DST, Govt. of India, for J. C. Bose Fellowship (JCB/2018/000047). S.P.S. further acknowledges the funding support from Intensification of Research in High Priority Areas (IRHPA) from the ANRF (IPA/2021/000096), and the Royal Society for Newton International Fellowship.

\item[Competing Interests] The authors declare that they have no competing financial interests.

\item[Data Availability] {The datasets analyzed in the current study are available from the corresponding authors upon reasonable request.}

\item[Correspondence]{All correspondence should be addressed to S.P. (email: shovon.pal@niser.ac.in)} or S.P.S. (email: satyaprasad@niser.ac.in)
\end{addendum}

\end{document}


\maketitle
\begin{affiliations}
\item School of Physical Sciences, National Institute of Science Education and Research, Jatni, 752 050 Odisha, India
\item Department of Electrophysics, National Yang Ming Chiao Tung University, Hsinchu 300, Taiwan
\item Homi Bhabha National Institute, Training School Complex, Anushakti Nagar, 400 094 Mumbai, India
\item Nanoelectronic and Device Physics Lab, School of Physical Sciences, National Institute of Science Education and Research, Jatni, 752 050, Odisha, India.
\item School of Chemical Sciences, National Institute of Science Education and Research, Jatni, 752 050 Odisha, India
\item Center for Interdisciplinary Sciences, National Institute of Science Education and Research, Jatni, 752 050 Odisha, India
\end{affiliations}

\vspace{-20pt}

\begin{flushleft}
(\today)
\end{flushleft}

\begin{abstract}
This supporting information contains further details on the sample characterization. The content is sectioned as follows:\\
\textbf{S1. Thickness measurements}\\
\textbf{S2. Bandgap evaluation and photoluminescence spectrum}\\
\textbf{S3. Tauc plot analysis}\\
\textbf{S4. Scanning Electron Microscopy}\\
\textbf{S5. Open-aperture (OA) Z-scan analysis}\\
\textbf{S6. Closed-aperture (CA) Z-scan analysis}\\
\textbf{S7. Transient absorption and phonons}\\
\textbf{S8. Literature comparison on the nonlinear optical coefficient}
\end{abstract}

\maketitle

\section{Thickness measurements}
\begin{figure}[b!]
    \centering
    \includegraphics[width=0.5\linewidth]{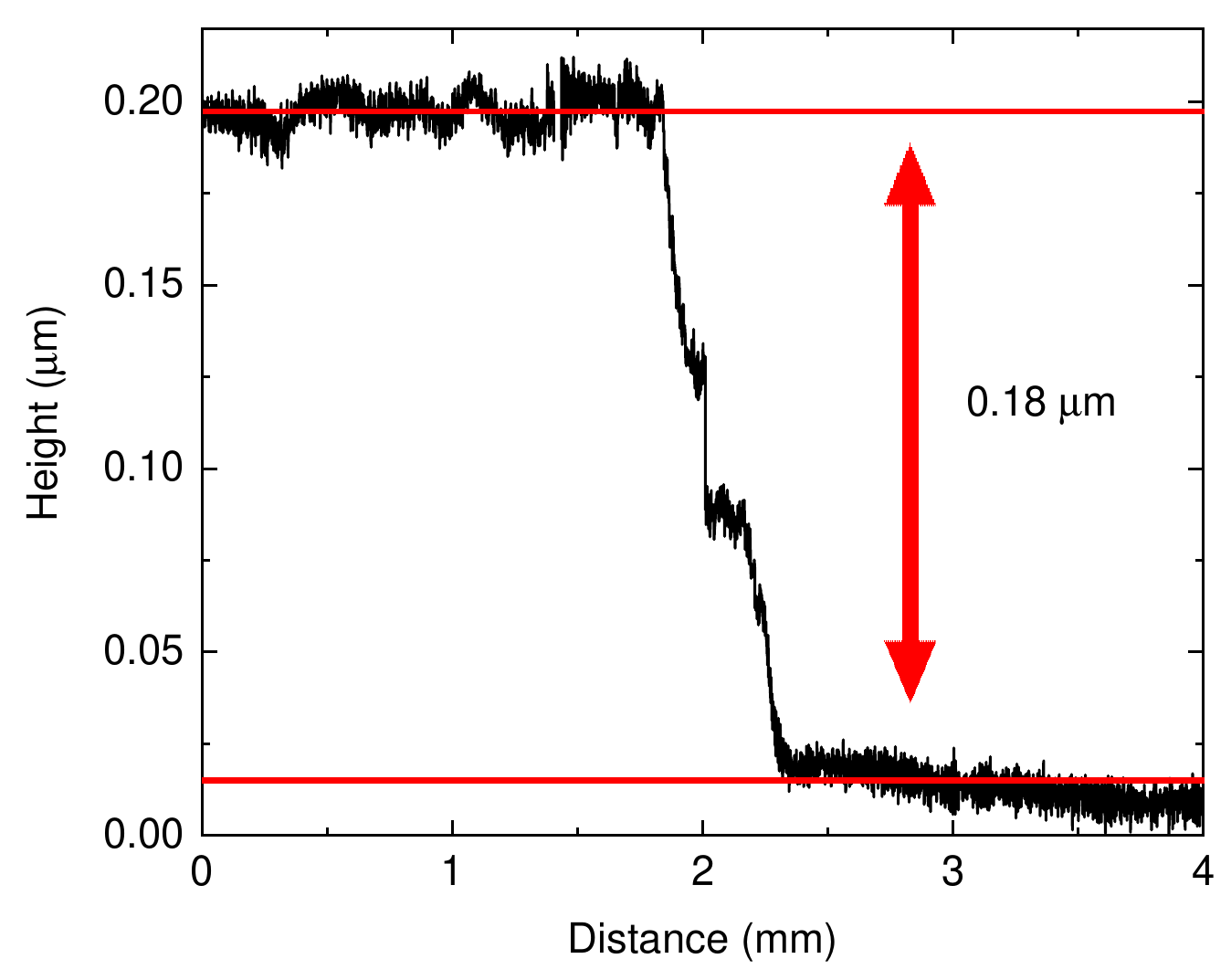}
    \vspace{-20pt}
    \caption{Profilometer scan over a length of 4\,mm, showing a relatively smooth surface of the (BA)$_2$PbI$_4$ film with a thickness of 0.18\,$\mu\text{m}$.}
    \label{Sfig2}
\end{figure}
The thickness of the spin-coated sample is measured using a surface profilometer (Bruker DektakXT). Prior to the measurement, a step edge is created on the sample surface by mechanically removing a portion of the film after deposition. The measurements are carried out by scanning the stylus across the step edge under controlled conditions. The stylus force, scan length, and scan speed are optimized to minimize surface damage while ensuring reliable height resolution. Multiple line scans are performed across different regions of the sample to account for thickness uniformity. The sample thickness is obtained from the height difference between the film surface and the bare substrate, as measured from the step profile. The final thickness is obtained by averaging the values measured across multiple scans, yielding 0.18\,$\mu\text{m}$. The associated uncertainty is estimated from the standard deviations of the data points in two regions of the step profile: the upper plateau (0.005\,$\mu\text{m}$) and the lower plateau (0.005\,$\mu\text{m}$). These two contributions were then combined to obtain the total uncertainty (0.01\,$\mu\text{m}$).

\section{Photoluminescence Characterization}
To investigate the optical properties and confirm the formation of the 2D perovskite phase, steady-state photoluminescence (PL) spectra of the (BA)$_2$PbI$_4$ films are acquired using an Edinburgh FS5 spectrofluorometer. Thin films are fabricated on $z$-quartz substrates by spin coating, as detailed in the methods section of the manuscript. Under an excitation wavelength ($\lambda_{\rm ex}$) of 400\,nm, the material exhibits a distinct, sharp emission peak centered at 540\,nm ($\approx 2.3$\,eV) as shown in Figure~\ref{Sfig2}a. The observed photoluminescence spectra show a Stokes shift of 76\,meV relative to the absorption band edge (see Figure~\ref{Sfig2}b for the bandgap evaluation from the Tauc plot). The magnitude of this shift is characteristic of the (BA)$_2$PbI$_4$ 2D structure and arises from the relaxation of charge carriers to the band edge before recombination. Importantly, the emission characteristics, specifically the strong excitonic features, align well with previously reported data for butylammonium-based lead iodide perovskites~\cite{Babaian2025JMCC}, distinguishing the 2D phase from its 3D counterparts. The spectral profile suggests that the luminescence originates primarily from the direct free-exciton recombination rather than defect-mediated transitions, thereby confirming the successful formation of the phase-pure (BA)$_2$PbI$_4$ perovskite.

\section{Tauc plot analysis}
\begin{figure}[t!]
    \centering
    \includegraphics[width=\linewidth]{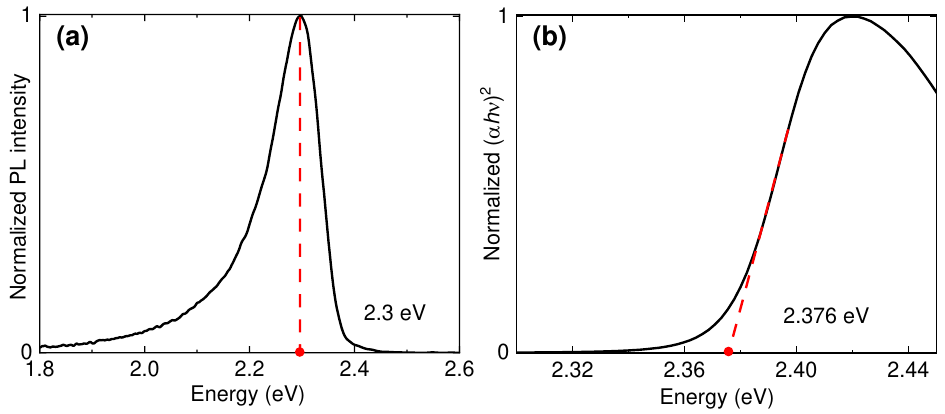}
    \caption{{\bf(a)} Photoluminescence (PL) spectrum of the (BA)$_2$PbI$_4$ film, which exhibits a sharp excitonic emission peak centered at 2.3\,eV. {\bf(b)} The Tauc plot of the (BA)$_2$PbI$_4$ film obtained from the UV-Visible absorption measurements, showing a linear region near the absorption edge. Extrapolation of the linear region indicates a direct optical bandgap of 2.376\,eV.}
    \label{Sfig2}
\end{figure}
UV-Visible absorption spectroscopy is employed to determine the optical band gap of (BA)$_2$PbI$_4$ film. The absorption spectrum is measured over the visible wavelength range ($400-800$\,nm). Background correction is performed by taking the substrate (on which the material is spin-coated) as a reference. The absorption coefficient ($\alpha$) is calculated from the measured absorbance ($A$) using the relation,
\begin{equation}
    \alpha = \frac{2.303\times A}{d} 
\end{equation}
where $d$ is the thickness of the film. Since (BA)$_2$PbI$_4$ is a direct band-gap semiconductor, the optical band gap is estimated using the Tauc relation~\cite{Tauc1966} for a direct allowed transition:
\begin{equation}
    (\alpha h \nu)^2 = B (h\nu - E_{\rm g}),
\end{equation}
where $B$ is a proportionality constant, $h\nu$ is the photon energy of the incident light, and $E_{\rm g}$ is the optical band gap of the sample. The Tauc plot of $(\alpha h \nu)^2$ vs Energy ($h\nu$) of (BA)$_2$PbI$_4$ film  is shown in Figure~\ref{Sfig2}b. The linear region near the absorption edge is extrapolated to $(\alpha h \nu)^2$ = 0. The intercept on the Energy axis ($h\nu = 2.376$~eV) determines the direct optical band gap of the (BA)$_2$PbI$_4$ thin film.

\section{Scanning Electron Microscopy}
To investigate the surface morphology and film quality of the deposited (BA)$_2$PbI$_4$ thin films, scanning electron microscopy (SEM) measurements are carried out using a Merlin Compact field-emission SEM (FESEM) equipped with a GEMINI-I electron column. To minimize possible electron beam-induced damage to the soft perovskite film, the images are recorded at a low accelerating voltage of 3\,kV with a working distance of 6\,mm. Notably, under these conditions, no observable beam damage is detected on the thin film. The SEM micrograph reveals a continuous, densely packed polycrystalline morphology with well-defined grains, indicating good film formation on the substrate with uniform surface coverage and high-quality, desirable for reliable optical characterization.

\begin{figure}[ht!]
    \centering
    \includegraphics[width=0.5\linewidth]{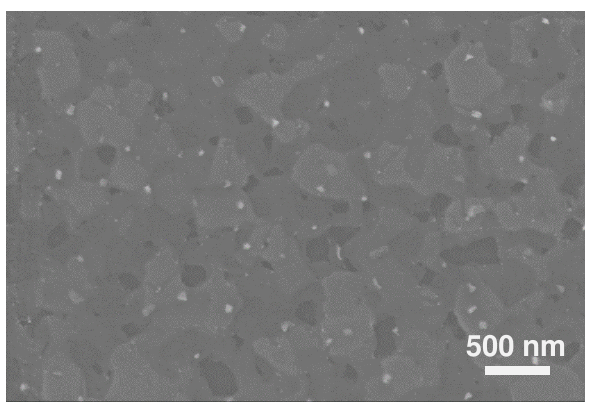}
    \caption{SEM micrograph of (BA)$_2$PbI$_4$ thin film. The image reveals a relatively uniform, nanostructured, densely packed grain distribution across the surface. No large cracks or pinholes are observed, indicating good surface coverage and film continuity. The scale bar corresponds to 500 nm.}
    \label{Sfig4}
\end{figure}

\section{Open-aperture (OA) Z-scan analysis} 
The propagation of a laser beam in an optically thin medium can be obtained using the following equation~\cite{Bahae1990IEEE}:
\begin{equation}
    \frac{dI(Z)}{dZ}=-\alpha_0 I(Z)-\sum_{n=2}\beta_n I(Z)^n.
    \label{Eq1}
\end{equation}
Here, $\alpha_0$ and $\beta_n$ are the linear, $n$-photon absorption coefficients, respectively. The position-dependent intensity $I(Z)$ can be obtained using the relation $I(Z) = I_0/(1+Z^2/Z_0^2)$, where $I_0$ is the on-axis peak intensity, and $Z_0$ is the Rayleigh length. In the case of the two-photon absorption, the solution of Eq.~\ref{Eq1} can be written as
\begin{equation}
    T_{\rm OA}(Z)=\frac{1}{\sqrt{\pi}\,(\beta_{2} I_{0} L_{\rm eff})/(1+Z^2/Z_0^2)} \int \ln\!\left[1+\frac{\beta_{2} I_{0} L_{\rm eff}}{1+Z^2/Z_0^2}\,e^{-x^{2}}\right]\,dx,
    \label{Eq2}
\end{equation}
where $T_{\rm OA}(Z)$ is the normalized open-aperture transmittance and $L_{\rm eff}=\left[1-\exp\!\left[(1-n)\alpha_0 L\right]\right]/(n-1)\alpha_0$ is the effective path length of the medium traversed by the interacting light pulse. $L$ is the length of the medium. Eq.~\ref{Eq2} can be further simplified as
\begin{equation}
    \ln\!\left|1-T_{\rm OA}(Z)\right|=\ln\!\left[I(Z)\right]+C.
    \label{Eq3}
\end{equation}

\begin{table}
    \centering
    \caption{Nonlinear absorption coefficients at different pump intensities.\\\\}
    \begin{tabular}{ccc}
    \hline \hline
    $I_0$ (GW/cm$^2$) & $-\beta \times 10^4$ (cm/GW) & $\gamma \times 10^4$ (cm$^3$/GW$^2$)\\
    \hline \hline
    18  & 69.8  & -   \\
    36  & 110.1 & -   \\
    54  & 157.2 & -   \\
    72  & 196.0 & -   \\
    81  & 280.2 & -   \\
    90  & 628.8 & 12.3\\
    95  & 388.7 & 16.1\\
    100 & -     & 18.3\\
    108 & -     & 20.5\\
    125 & -     & 23.8\\
    \hline \hline
    \end{tabular}
\end{table}

\noindent In the same fashion, for the case of the three-photon absorption, the normalized open-aperture transmittance can be expressed as
\begin{equation}
    \ln\!\left|1-T_{\rm OA}(Z)\right|=2\ln\!\left[I(Z)\right]+C'.
    \label{Eq4}
\end{equation}
Here, $C$ and $C'$ are integration constants. Evidently, in the case of the OA Z-scan, $\ln\!\left|1-T_{\rm OA}(Z)\right|$ varies linearly with $\ln\!\left[I(Z)\right]$ with a slope of $n$ in case of the $(n+1)$-photon absorption. This relation is used to identify the order of the nonlinear optical response~\cite{Steinhoff2017NatComm, Li2018SciRep}. A more generalized solution of Eq.~\ref{Eq1} for determining the nonlinear absorption coefficient  takes the following form~\cite{Boyd2008APB, Gaur2018JOSA}
\begin{equation}
    T_{\rm OA}(Z)=\sum_{n=0}^{\infty}\frac{(-1)^n}{(n+1)^{3/2}} \left[\frac{\beta I_0\,L_{\mathrm{eff}}}{1+Z^2/Z_0^2}\right]^n.
    \label{Eq5}
\end{equation}
If we now restrict Eq.~\ref{Eq1} to three-photon absorption and include saturation of two-photon absorption, it takes the form~\cite{Gaur2018JOSA}
\begin{equation}
    \frac{dI}{dZ}=-\alpha_0 I-\frac{\beta I^2}{1+\dfrac{I^2}{I_s^2}}-\gamma I^3,
    \label{Eq6}
\end{equation}
where $\beta$ and $\gamma$ represent the two-photon absorption and three-photon absorption coefficients, respectively. $I_s$ represents the saturation intensity for two-photon absorption. Under such a condition, the nonlinear absorption coefficient of the medium can be written as~\cite{Boyd2008APB, Gaur2018JOSA}
\begin{equation}
    \alpha(I)=\alpha+\frac{\beta I}{1+\dfrac{I^2}{I_s^2}}+\gamma I^{2}.
    \label{Eq7}
\end{equation}
We can extract these nonlinear optical coefficients by fitting the OA Z-scan data with Eq.~\ref{Eq5} after replacing the $\frac{\beta I_0}{1 + Z^2/Z_0^2}$ term with Eq.~\ref{Eq7}. The nonlinear coefficients for different fluences in our experiments are summarized in Table~1.

\section{Closed-aperture Z-scan analysis} 
In the case of CA Z-scan measurements, one can isolate the pure refraction contribution using the relation T$_{\rm Refrac}(Z)$  = T$_{\rm CA}$/T$_{\rm OA}$. For a Gaussian beam traveling in the $Z$-direction, the nonlinear phase acquired by the beam can be written as~\cite{Bahae1990IEEE}
\begin{equation}
    \frac{d\Delta\phi}{dZ}=-\,\Delta n(I)\,k,
    \label{Eq8}
\end{equation}
from where the solution of the phase shift is 
\begin{equation}
    \Delta\phi(Z,r,t)=\Delta\phi_0(Z,t)\,\exp\!\left(-\frac{2r^2}{\omega^2}\right).
    \label{Eq9}
\end{equation}
Here, $\Delta\phi_0(Z,t)=\Delta\phi_0(t)\left(1+\dfrac{Z^2}{Z_0^2}\right)^{-1}$ is the on-axis position-dependent phase change, $r$ is the radial direction, and $\Delta\phi_{0}(t)=kn_{2}I_{0}L_{\rm eff}$ is the on-axis phase shift at the focus. The complex electric field used to excite the material carries forward the nonlinear phase and takes the form~\cite{Bahae1990IEEE, Fallah2025JOSA}
\begin{equation}
    E_{\rm exc}=E_{\rm incident}(Z,r,t)\,e^{-\frac{\alpha_0 L}{2} + i\,\Delta\phi(Z,r,t)}.
    \label{Eq10}
\end{equation}
Assuming a circularly symmetric Gaussian beam propagating in the positive $Z$-direction, the electric field can be expressed as
\begin{equation}
    E(r,Z,t)=E(t)\,\frac{\omega_0}{\omega(Z)}\,e^{\,i\,\!\left(kZ+\phi(Z)-\frac{k r^2}{2R(Z)}\right)-\frac{r^2}{2\omega^2(Z)}}.
    \label{Eq11}
\end{equation}
Here, $E(t)$ is the temporal profile of the beam at the focal point, $\omega(Z)=\omega_0\sqrt{1+\frac{Z^2}{Z_0^2}}$ is the position-dependent beam radius. $\omega_0$ is the beam waist, $\phi(Z)=\tan^{-1}\!\left(\frac{Z}{Z_0}\right)$ and $R(Z)=Z+\frac{Z_0^2}{Z}$ is the radius of curvature. Using Eq.~\ref{Eq10} and~\ref{Eq11}, we can calculate the electric field at the aperture $E_a (\rho,z,t)$ as
\begin{equation}
    E_a(\rho,Z,t)=\frac{e^{ik\left(d+\frac{\rho^2}{2d}\right)}}{i d \lambda}\int_0^{\infty} J_0\!\left(\frac{k r \rho}{d}\right)E_{\rm exc}(Z,r,t)\, 2\pi r\, dr,
    \label{Eq12}
\end{equation}
where $d$ is the distance between the sample and the aperture, and $J_0(kr\rho/d)$ is the first-order Bessel function. We can further determine the normalized transmittance for CA Z-scan, i.e., $T_{\rm Refraction}(Z)$ as
\begin{equation}
    T_{\rm Refraction}(Z)=\frac{\displaystyle \int_0^a\left| E_a(\rho,Z,\Delta\theta) \right|^2\,\rho\, d\rho} {\displaystyle \int_0^a \left| E_a(\rho,Z,\Delta\theta=0) \right|^2\,\rho\, d\rho}.
    \label{Eq13}
\end{equation}
Here, $a$ is the radius of the aperture. Further, under far-field conditions $(d>>Z_0)$ and small nonlinear optical phase change Eq.~\ref{Eq13} can be simplified to~\cite{Bahae1990IEEE, Fallah2025JOSA}
\begin{equation}
    T_{\rm Refraction}(Z)=1+\frac{4\left(Z/Z_{0}\right)}{\left[\left(Z/Z_0\right)^2+9\right]\left[\left(Z/Z_0\right)^2+1\right]}\,\Delta\phi_0^{(3)}
    \label{Eq14}
\end{equation}
Finally, the coefficient of nonlinear refraction ($n_2$) is determined by using the relation $\Delta\phi^{(3)}_0=2\pi n_2I_0L_{\rm eff}/\lambda$.

\section{Transient absorption and phonons}
\begin{figure}[ht!]
    \centering
    \includegraphics[width=1.0\linewidth]{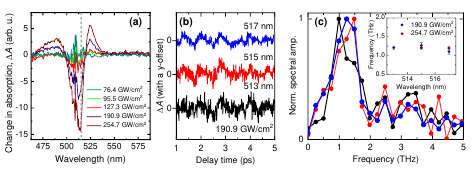}
    \caption{{\bf(a)} Transient absorption (TA) spectra of (BA)$_2$PbI$_4$ thin film as a function of probe wavelength when pumped at 1030\,nm. The vertical gray-dashed line indicates the position where the transient analysis is carried out in Figure~4a of the main manuscript. {\bf(b)} The background-subtracted TA signal at three different probe wavelengths for the pump intensity of 190.9\,GW/cm$^2$. These oscillations are attributed to the phonon mode of (BA)$_2$PbI$_4$. {\bf(c)} The corresponding FFT spectra of the time transients at different probe wavelengths shown in (b), revealing the phonon frequency at 1.2\,THz. The inset shows that the phonon mode frequency is independent of probe wavelength and varies minimally with higher pump intensity.}
    \label{Sfig5}
\end{figure}
It can be noted that for higher pump fluences, the transient absorption data reveal characteristic (and pronounced) oscillations as the signal (i.e., $\Delta A$) relaxes, shown in Figure~4a of the main manuscript for $I_0=190.9$\,GW/cm$^2$.   These oscillations persist for approximately $5-6$\,ps. Figure~\ref{Sfig5}b shows these exemplary oscillations after subtracting the slow-varying background. A fast Fourier transform of these extracted oscillatory signals reveals a peak around 1.2\,THz, see Figure~\ref{Sfig5}c. This oscillation is attributed to the coherent Pb-I-Pb lattice vibration, which is directly captured in the transient absorption signal under high-pump-intensity excitation. In contrast, at lower pump intensity, the oscillatory feature could not be resolved clearly due to low amplitudes of the phonon oscillations. Also note that the phonon mode remains unchanged across different probe wavelengths, indicating that the oscillatory behavior arises from intrinsic lattice vibrations rather than probe-energy-dependent electronic effects. All parameters obtained from the fitting of the transient absorption data are listed in Table~2.

\begin{table}
    \centering
    \caption{Parameters obtained from the bi-exponential fitting of the transient absorption data in Figure~4a of the main manuscript.\\}
    \begin{tabular}{ccccc}
    \hline \hline
    $I_0$ (GW/cm$^2$) & $A_1\times10^{-3}$ (arb. u.) & $A_2\times10^{-3}$ (arb. u.) & $t_1$ (ps) & $t_2$ (ps) \\
    \hline \hline
    $76.4$  & $-(0.27 \pm 0.07)$  & $0.55 \pm 0.027$ & $0.18 \pm 0.07$ & $18.96 \pm 3.33$ \\
    $95.5$ &  $-(0.56 \pm 0.06)$  & $0.39 \pm 0.02$ & $0.16 \pm 0.03$ & $34.65 \pm 6.59$ \\
    $127.3$  & $-(0.87 \pm 0.05)$  & $-(0.47 \pm 0.01)$ & $0.21 \pm 0.03$ & $500.0 \pm 71.20$ \\
    $190.9$  & $-(1.10 \pm 0.07)$ & $-(3.78 \pm 0.09)$ & $0.20 \pm 0.02$ & $343.09 \pm 12.67$ \\
    $254.7$  & $-(3.19 \pm 0.12)$ & $-(6.91 \pm 0.13)$ & $0.17 \pm 0.02$ & $427.55 \pm 13.03$ \\
    \hline \hline
    \end{tabular}
\end{table}

\section{Literature comparison on the nonlinear optical coefficient}
In Table~3 presented below, we compare our nonlinear absorption coefficient with reported values for related two-dimensional perovskites and other representative nonlinear optical materials. The table includes the material, material type, excitation wavelength, pulse width, and nonlinear absorption coefficient. We have selected literature reporting saturable absorption under excitation conditions similar to those used in the present work.

The comparison shows that the nonlinear absorption coefficient measured for our polycrystalline (BA)$_2$PbI$_4$ film at 1030\,nm with 370\,fs pulses is among the largest values reported for related perovskite systems.

\begin{table}
    \centering
    \caption{Comparison of the nonlinear absorption coefficient ($\beta$) reported for various material systems under excitation conditions similar to those used in the present work.\\}
    \label{tab:beta_comparison}
    \begin{tabular}{ccccc}
    \hline \hline
    Material System & Material Class & $\lambda$ (nm) & $\tau$ (fs) & $|\beta|$ (cm/GW) \\
    \hline \hline
    Graphene~\cite{Kumar2009APL} & 2D Carbon & 790 & 100 & 90\\
    72--74 L MoS$_2$~\cite{Zhang2015ACSN} & 2D TMDCs & 1030 & 340 & 250\\
    18--20 L WS$_2$~\cite{Zhang2015ACSN} & 2D TMDCs & 800 & 40 & 397\\
    Black Phosphorus~\cite{Lu2015OE} & 2D Semiconductor & 800 & 140 & $6.17 \times 10^{-3}$\\
    Bi$_2$Se$_3$~\cite{Wang2016SR} & Topological Insulator & 800 & 100 & $6.5 \times 10^{4}$\\
    Bi$_2$Te$_3$~\cite{Wang2016SR} & Topological Insulator & 800 & 100 & $2.1 \times 10^{4}$\\
    CH$_3$NH$_3$PbI$_3$~\cite{Zhang2016ACSP} & 3D Perovskite Thin Film & 1064 & 40 & $2.25 \times 10^{3}$\\
    CH$_3$NH$_3$PbI$_{3-x}$Cl$_x$~\cite{Zhang2016ACSP} & 3D Perovskite Thin Film & 1064 & 40 & $2.03 \times 10^{3}$\\
    Violet Phosphorus~\cite{Guo2024Mat} & 2D Phosphorus Nanosheets & 1030 & 100 & $2.18 \times 10^{7}$\\
    (PEA)$_2$Cs$_4$Pb$_5$Br$_{16}$~\cite{Mushtaq2022MA} & Quasi-2D RP Perovskite Film & 800 & 57 & 13.83\\
    Cs$_2$NaBiI$_6$~\cite{Singh2022PCCP} & Inorganic Double Perovskite Film & 1030 & 370 & $6.23 \times 10^{3}$\\
    Cs$_2$KBiI$_6$~\cite{Singh2022PCCP} & Inorganic Double Perovskite Film & 1030 & 370 & $1.14 \times 10^{3}$\\
    (BA)$_2$PbI$_4$ film~\cite{Liang2020OExp} & 2D RP Perovskite & 500 &  100 &  $ 2.4 \times 10^{5}$\\
    {\bf (BA)$_2$PbI$_4$ film} $^{\text{ this work}}$ & {\bf 2D RP Perovskite} & {\bf 1030} & {\bf 370} &  ${\bf 6.29} \times {\bf 10^{6}}$\\
    \hline \hline
    \end{tabular}
\end{table}